# A Novel Class-F 2.45/5.8 GHz Dual-Band Rectifier for Wireless Power Transmission

Siyi Xiao, *Graduate Student Member, IEEE*, Changjun Liu, *Senior Member, IEEE*, Haoming He, *Graduate Student Member, IEEE*, Yuhao Feng, and Wenquan Che, *Fellow, IEEE*

*Abstract*—This letter proposes a high-efficiency dual-band class-F rectifier for wireless power transmission (WPT). The rectifier comprises a dual-band harmonic termination network, a dual-band matching network, a single Schottky diode, and a dc pass filter. A theoretical analysis of the harmonic termination network is performed to improve the rectifying efficiency. The network exhibits good class-F operation by controlling the second and third harmonics in dual bands. A rectifier operating at 2.45 and 5.8 GHz was designed, fabricated, and measured for validation. The measurements show maximum RF-dc conversion efficiencies of 74.9% and 61.9% with 200 and 500 Ω loads at 2.45 and 5.8 GHz, respectively. The proposed rectifier achieves dual-band harmonic control with high efficiency.

## I. INTRODUCTION

**W**ITH the development of wireless electronic devices, wireless power transmission (WPT) has become more prevalent [1], [2], [3]. Microwave rectifiers, which convert radio frequency (RF) power into dc power [4], [5], are essential components in WPT systems. A dual-band rectifier that can harvest microwave energy in two bands is more versatile than a single-band rectifier [6], [7], [8], [9], [10], [11], [12]. However, designing a high-efficiency dual-band rectifier is challenging since a rectifying diode's input impedance varies nonlinearly with the operating frequencies and input power levels [13], [14], [15].

Recently, research groups have developed several rectifiers with dual-band matching networks to solve those issues, such as L-type [7], [16], T-type [9], [17], multistub [18], [19], [20], *LC* networks, and so on. In [18], a dual-band impedance compression network was successfully proposed for the first time to reduce the variation of the input impedance of rectifiers with input power. Two sub-rectifiers with four diodes and two dc loads were used in the dual-band design.

Manuscript received 18 July 2023; revised 29 August 2023 and 15 October 2023; accepted 29 October 2023. Date of publication 15 November 2023; date of current version 9 January 2024. This work was supported in part by NSFC under Grant U22A2015, Grant 62071316, and Grant 61931009. (*Corresponding authors: Changjun Liu; Wenquan Che.*)

Siyi Xiao, Changjun Liu, and Haoming He are with the School of Electronics and Information Engineering, Sichuan University, Chengdu 610064, China, and also with the Yibin Industrial Technology Research Institute, Sichuan University, Yibin 644001, China (e-mail: cjliu@ieee.org).

Yuhao Feng is with the School of Electronic Science and Engineering, University of Electronic Science and Technology, Chengdu 611731, China.

Wenquan Che is with the School of Electronic and Information Engineering, South China University of Technology, Guangzhou 510641, China (e-mail: eewqche@scut.edu.cn).

Color versions of one or more figures in this letter are available at https://doi.org/10.1109/LMWT.2023.3330322.

Digital Object Identifier 10.1109/LMWT.2023.3330322

Moreover, several research groups successfully applied novel harmonic controlling networks to improve rectifying efficiency [21], [22]. Liu [21] used a series band-stop structure to achieve harmonic recycling, resulting in 80.9% efficiency at 2.45 GHz, which is also applied to a voltage-doubling rectifier in [22] as well. A few rectifiers, such as class-C, class-F, and class-F$^{-1}$ rectifiers, are equipped with typical harmonic controlling methods reported in [23], [24], [25], and [26], respectively. In [23], an analytical model was derived to calculate rectifying efficiency in class-F rectifiers based on the diode voltage and current waveforms in the time domain. Therefore, those studies focus more on single-band rectifiers and have limitations in dual-band applications.

This letter proposes a high-efficiency dual-band rectifier based on a novel class-F load. The harmonic termination network controlling the second and third harmonics of dual bands, which shapes the voltage and current waveforms into a square and half sine waves, respectively, improves the dual-band rectifying efficiency. In addition, a theoretical analysis of the network is performed, playing a crucial role in rectifier design. Using the proposed design method, a high-efficiency facilitated and compact dual-band rectifier is achieved with rectifying efficiencies of 74.9% and 61.9% at 2.45 and 5.8 GHz, respectively.

## II. PRINCIPLE AND DESIGN METHOD

A schematic of the proposed dual-band rectifier is illustrated in Fig. 1. The rectifier comprises a dual-band harmonic termination network, a dual-band impedance matching network, a Schottky diode (HSMS286F), a dc-pass filter, and a dc load ($R_L$).

The second and third harmonics usually are stronger than high-order harmonics, which should be recycled and converted into dc again to improve the rectifying efficiency.

Fig. 2 depicts the design of the harmonic termination network. The lines $TL_1$, $TL_2$, and $TL_3$ are designed to control the third harmonics, and the lines, $TL_1$, $TL_2$, $TL_3$, $TL_4$, $TL_5$, and $TL_6$ are designed to control the second harmonics, which present with zero impedance at second harmonics and infinite impedance at third harmonics to the diode at plane $P_1$.

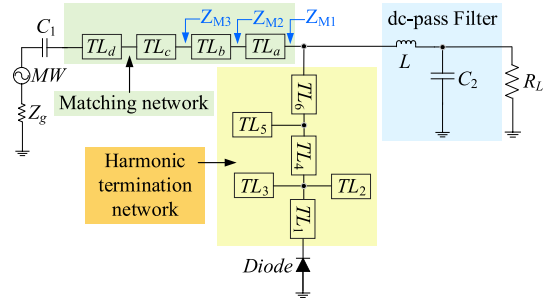

Fig. 1. Schematic of the proposed dual-band rectifier.

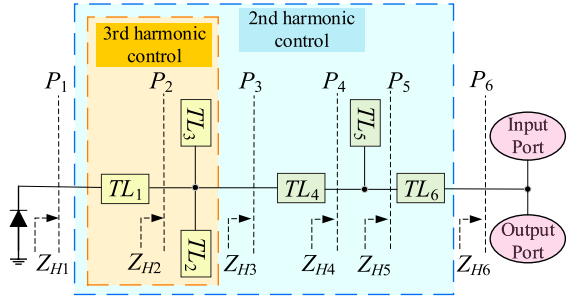

Fig. 2. Schematic of the proposed harmonic termination network.

TABLE I
PARAMETERS OF HARMONIC TERMINATION NETWORK

| $Z_{TL1}$ (Ω) | $Z_{TL2}$ (Ω) | $Z_{TL3}$ (Ω) | $Z_{TL4}$ (Ω) | $Z_{TL5}$ (Ω) | $Z_{TL6}$ (Ω) |
|---|---|---|---|---|---|
| 52 | 68 | 64 | 70 | 70 | 90 |

| $\theta_{TL1}$ (deg) | $\theta_{TL2}$ (deg) | $\theta_{TL3}$ (deg) | $\theta_{TL4}$ (deg) | $\theta_{TL5}$ (deg) | $\theta_{TL6}$ (deg) |
|---|---|---|---|---|---|
| 20 | 30 | 12.7 | 53 | 45 | 9 |

The ratio of two fundamental frequencies is defined as follows:

$$f_2/f_1 = k. \tag{1}$$

The ratios of the different fundamental and harmonic frequencies $f_i$ ($i = 1, 2, 3, 4, 5, 6$) to the fundamental frequency $f_1$ are defined as follows:

$$k_i = \begin{cases} 1, & i = 1@\,f_1 \\ k, & i = 2@\,f_2 \\ 2, & i = 3@\,f_3 = 2f_1 \\ 2k, & i = 4@\,f_4 = 2f_2 \\ 3, & i = 5@\,f_5 = 3f_1 \\ 3k, & i = 6@\,f_6 = 3f_2. \end{cases} \tag{2}$$

First, the electrical lengths of TL2 and TL3 are known as follows:

$$\theta_{TL2}@\,3f_1 = 90°, \quad \theta_{TL3}@\,3f_2 = 90°. \tag{3}$$

When the short circuit conditions are satisfied at plane $P_2$ in Fig. 2, the transmission line TL1 is used to control the input impedances at $3f_1$ and $3f_2$, and the impedances $Z_{H1}(3f_1)$, $Z_{H1}(3f_2)$ are as follows:

$$\begin{aligned} Z_{H1}(3f_1) &= jZ_{TL1}\tan(3\theta_{TL1}) \\ Z_{H1}(3f_2) &= jZ_{TL1}\tan(3k\theta_{TL1}). \end{aligned} \tag{4}$$

From (3) and (4), the electrical lengths $\theta_{TL1}$, $\theta_{TL2}$, and $\theta_{TL3}$ are calculated. The characteristic impedances $Z_{TL1}$, $Z_{TL2}$, and $Z_{TL3}$ are free parameters whose values will affect the impedance $Z_{H2}$ at plane $P_2$.

According to the transmission line input impedance formula, the impedance $Z_{H1}(f_i)$ ($i = 1, 2, 3, 4$) at plane $P_1$ is calculated by the following equation:

$$Z_{H1}(f_i) = Z_{TL1} \times \frac{Z_{H2}(f_i) + jZ_{TL1}\tan(k_i\theta_{TL1})}{Z_{TL1} + jZ_{H2}(f_i)\tan(k_i\theta_{TL1})}. \tag{5}$$

By rearranging (5), the impedance $Z_{H2}(f_i)$ ($i = 1, 2, 3, 4$) at plane $P_2$ is easily expressed as follows:

$$Z_{H2}(f_i) = Z_{TL1} \times \frac{Z_{H1}(f_i) - jZ_{TL1}\tan(k_i\theta_{TL1})}{Z_{TL1} - jZ_{H1}(f_i)\tan(k_i\theta_{TL1})}. \tag{6}$$

The impedance $Z_{H2}(f_i)$ ($i = 1, 2, 3, 4$) can also be obtained by the following equation:

$$\begin{aligned} Z_{H2}(f_i) \\ = Z_{H3}(f_i) \parallel (-jZ_{TL2}\cot(k_i\theta_{TL2})) \parallel (-jZ_{TL3}\cot(k_i\theta_{TL3})). \end{aligned} \tag{7}$$

The impedances $Z_{H3}(f_i)$ ($i = 1, 2, 3, 4$) at plane $P_3$ are obtained by combining (6) and (7).

Furthermore, the electrical length of TL5 is expressed as follows:

$$\theta_{TL5}@\,2f_1 = 90°. \tag{8}$$

Since the short circuit condition at $2f_1$ is satisfied at plane $P_4$, when the characteristic impedance $Z_{TL4}$ is a free parameter, the electrical length $\theta_{TL4}$ is calculated by the following equation:

$$\theta_{TL4} = 0.5 \times \arctan(Z_{H3}(2f_1)/jZ_{TL4}). \tag{9}$$

Similarly, the impedance $Z_{H4}(2f_2)$ at plane $P_4$ is obtained by the following equation:

$$\begin{aligned} Z_{H4}(2f_2) &= Z_{TL4}\frac{Z_{H3}(2f_2) - jZ_{TL4}\tan(2k\theta_{TL4})}{Z_{TL4} - jZ_{H3}(2f_2)\tan(2k\theta_{TL4})} \\ Z_{H4}(2f_2) &= Z_{H5}(2f_2) \parallel (-jZ_{TL5}\cot(2k\theta_{TL5})). \end{aligned} \tag{10}$$

From (10), the impedances $Z_{H5}(2f_2)$ at plane $P_5$ are calculated. The transmission line TL6 is used to control $2f_2$. The TL6 is followed by a fundamental frequency matching network and a dc filter, and the impedance of harmonics at plane $P_6$ does not affect the design of the matching network and dc filter.

Then, when the second harmonic impedance of $f_2$ at plane $P_6$ is infinite, the impedance $Z_{H5}(2f_2)$ at plane $P_5$ is as follows:

$$Z_{H5}(2f_2) = -jZ_{TL6}\cot(2k\theta_{TL6}). \tag{11}$$

The characteristic impedance $Z_{TL6}$ is a free parameter. The electrical length $\theta_{TL6}$ is calculated according to (11), once $Z_{TL6}$ is chosen. Since $Z_{H5}(2f_2)$ is a fixed value and is not affected by the line TL6, this assumption does not affect the class-F mechanism. Therefore, as long as (11) has a solution, the assumption is true.

As a result, all electrical lengths are calculated and all the impedances are free design parameters, giving more design space for impedance matching at dual fundamental frequencies. The parameters are given in Table I.

In particular, when $k = 1$, it is a single-band rectifier, and the electrical lengths of TL1, TL2, TL3, TL4, and TL5, are as follows:

$$\begin{aligned} \theta_{TL1} &= 90°, \quad \theta_{TL2} = \theta_{TL3} = 30° \\ \theta_{TL4} &= 180°, \quad \theta_{TL5} = 45°. \end{aligned} \tag{12}$$

Therefore, when $k = 1$, the proposed harmonic termination network can be simplified into the single-band class-F harmonic termination network in literature [23], which extends its application to the dual-band rectifier.

Moreover, the rectifying efficiencies of ideal rectifiers with and without harmonic termination networks are both simulated. The rectifying efficiencies are improved by about 10% in dual bands with the harmonic termination networks in Fig. 3.

 

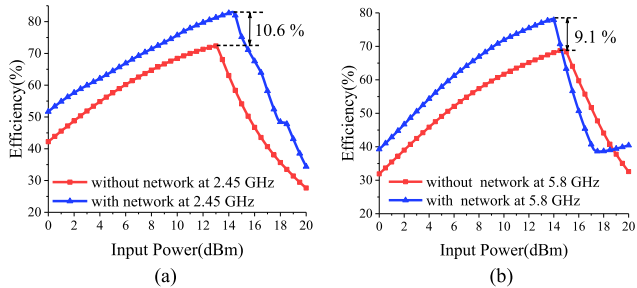

Fig. 3. Simulated the RF-dc conversion efficiencies of the ideal rectifiers with and without harmonic termination networks. (a) 2.45 GHz. (b) 5.8 GHz.

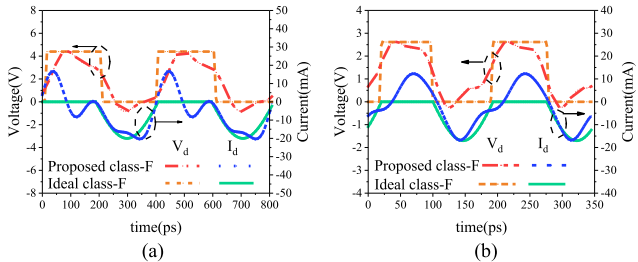

Fig. 4. Simulated diode voltage and current waveforms of the proposed rectifier and the ideal class-F rectifier. (a) 2.45 GHz. (b) 5.8 GHz.

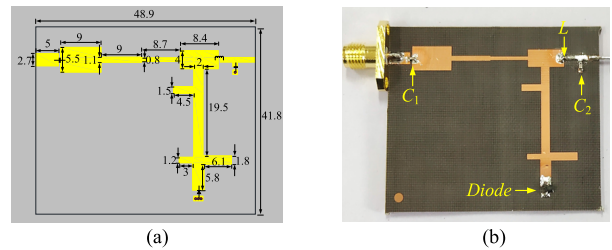

Fig. 5. (a) Layout of the proposed rectifier (unit: mm). (b) Fabricated rectifier.

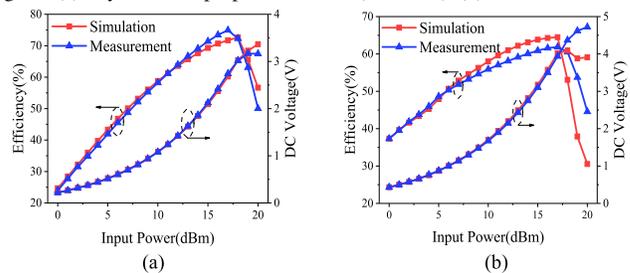

Fig. 6. Simulated and measured RF-dc conversion efficiencies and dc output voltages with respect to input power at (a) 2.45 GHz with $R_L = 200$ Ω and (b) 5.8 GHz with $R_L = 500$ Ω.

Fig. 4 shows the simulated time-domain voltage and current waveforms of the proposed rectifier and ideal class-F rectifier at 2.45 and 5.8 GHz, which means the proposed rectifier has a good class-F operation. The class-F mechanism shapes the current and voltage waveforms into square and half-sine waves, respectively. The overlap of voltage and current waveforms is minimized with the wave-shaping mechanism, so the rectifying efficiency is improved.

Four microstrip lines are used to match the different impedances in dual bands in Fig. 1 [27], [28]. The complex impedances $Z_{M1}(f_1)$ and $Z_{M1}(f_2)$ at two fundamental frequencies $f_1$ and $f_2$ are $R_1 + jX_1$ and $R_2 + jX_2$, respectively. First, a series microstrip line $\text{TL}_a$ transforms $Z_{M1}$ at dual frequencies into a pair of conjugate complex impedances $Z_{M2}$. Therefore, the complex impedances $Z_{M2}(f_1)$ and $Z_{M2}(f_2)$ at $f_1$ and $f_2$ are $R_3 + jX_3$, $R_3 - jX_3$, respectively. Then, a series microstrip line $\text{TL}_b$ tunes the pair of conjugate complex impedances $Z_{M2}$ into the same resistance $R_{M3}$. Finally, two series microstrip lines $\text{TL}_c$, and $\text{TL}_d$ are used to tune the resulting real impedance $R_{M3}$ to 50 Ω.

## III. Implementation and Measurement

A dual-band rectifier designed to operate at 2.45 and 5.8 GHz has been implemented and tested. The substrate of F4B-2 (polytetrafluoroethylene microfiber glass) was used with a relative dielectric constant of 2.65, thickness of 1 mm, and loss tangent of 0.002. A capacitor $C_1$ (30 pF) is applied as a dc-block, while the dc-pass filter consists of an inductor (13 nH) and a capacitor $C_2$ (100 pF). The rectifier dimensions are 48.9 × 41.8 mm, as shown in Fig. 5.

The RF power is generated using an Agilent E8267C microwave source. The output dc voltage is measured with a voltage meter, and a standard resistor box serves as the output dc load. The simulated and measured efficiencies of the proposed rectifier at 2.45 GHz with a 200 Ω load are shown in Fig. 6(a), which is in good agreement. At the input power of 17 dBm, the maximum measured RF-dc conversion efficiency at 2.45 GHz is 74.9%, with a peak output dc voltage of 2.74 V. At 5.8 GHz, Fig. 6(b) displays the RF-dc conversion efficiency of 61.9% at the input power of 17 dBm with a 500 Ω load, with a peak output dc voltage of 3.94 V.

The measured efficiency at the higher frequency is slightly lower than the simulation, which may be due to manufacturing tolerances and differences between the diode model and the real one in C-band. Moreover, a diode's characteristics, such as impedance and junction capacitance, are strongly dependent on frequency. Thus, the optimal dc load of a rectifier also varies with frequency, so the dc loads of two frequencies are different in the actual measurements.

Table II compares the performance of the proposed rectifier with previous publications. The proposed rectifier achieves a higher RF-dc conversion efficiency with only one diode and one dc load. It realizes efficient harmonic control at dual bands.

## IV. Conclusion

In this letter, a high-efficiency class-F dual-band rectifier is proposed. A novel harmonic termination network is designed to control second and third harmonics in both operating bands and applied to improve rectifying efficiency. The measured results show that maximum rectifying efficiencies reach 74.9% and 61.9% at 2.45 and 5.8 GHz, respectively. The proposed rectifier is suitable for WPT applications, such as charging wireless sensors in IoTs.

### TABLE II
### Comparison of Previously Reported Dual-Band Rectifiers

| Ref | Frequency (GHz) | Input Power (dBm) | Peak efficiency (%) | Diode (HSMS) | Electrical Size ($\lambda_g \times \lambda_g$) | Number of diodes and sub-rectifiers | Harmonic control |
|---|---|---|---|---|---|---|---|
| [6] | 0.915, 2.45 | 17 | 74.9, 71.2 | 286 | 0.73×0.26 | 4, 2 | no |
| [7] | 2.45, 5.8 | 11 | 64.8, 64.2 | 286 | 0.73×0.61 | 1, 1 | yes |
| [10] | 0.915, 2.45 | 30 | 66, 58 | 282, 285 | N/A | 4, 2 | no |
| [12] | 3.5, 5.8 | 0 | 51.8, 39.7 | 286 | 0.44×0.55 | 1, 1 | no |
| [16] | 0.915, 2.45 | 28 | 71, 67.7 | 282 | 0.22×0.08 | 2, 1 | no |
| [18] | 2.45, 5.8 | 18.1 | 76.2, 61 | 286 | N/A | 4, 2 | no |
| [19] | 2.45, 5.8 | 16 | 55.9, 55.4 | 286 | 0.95×0.5 | 1, 1 | no |
| [20] | 2.45, 5.8 | 8 | 57.6, 33.6 | 285 | 1.08×0.41 | 1, 1 | no |
| This work | 2.45, 5.8 | 17 | 74.9, 61.9 | 286 | 0.65×0.55 | 1, 1 | yes |

$\lambda_g$: the wavelength referring to the lower frequency of the dual bands.